\documentclass{iau} 

\usepackage{graphicx}
\usepackage{xcolor}
\usepackage[hyphens]{url}
\usepackage{hyperref}
\hypersetup{
     colorlinks  = true,
     linkcolor   = blue, 
     anchorcolor = BrickRed,
     citecolor   = blue,
     filecolor   = blue,
     menucolor   = blue,
     runcolor    = cyan,
     urlcolor    = blue
}
\usepackage[authoryear]{natbib}

\pubyear{2019}
\volume{355} 
\jname{The Realm of the Low-Surface-Brightness Universe}
\editors{D. Valls-Gabaud, I. Trujillo \& S. Okamoto, eds.}
\title[Baryon dominated UDGs] %% give here short title %%
{Baryon dominated ultra-diffuse galaxies\\ }

\author[Mancera Pi\~na, Fraternali, Adams \& Oosterloo]   %% give here short author list %%
{Pavel E. Mancera Pi\~na$^{1,2}$,
 Filippo Fraternali$^1$, Elizabeth A. K. Adams$^{2,1}$ and Tom Oosterloo$^{2,1}$ }

\affiliation{$^1$Kapteyn Astronomical Institute, University of Groningen, Landleven 12, 9747 AD, Groningen, The Netherlands \\[\affilskip]
$^2$ASTRON, Netherlands Institute for Radio Astronomy, Postbus 2, 7900 AA Dwingeloo, The Netherlands \\}
\begin{document}

\maketitle

\begin{abstract}
By means of interferometic 21-cm observations and a 3D kinematic modeling technique, we study the gas kinematics of six H\,{\sc i}--rich ultra-diffuse galaxies (UDGs). We derive robust circular velocities and baryonic masses, that allow us to study the position of our UDGs with respect to the baryonic Tully-Fisher relation (BTFR). Somewhat surprisingly, we find that these galaxies are strong outliers from the BTFR, rotating too slowly for their baryonic mass. Moreover, their position in the circular velocity--baryonic mass plane implies that they have a baryon fraction inside their virial radii compatible with the cosmological mean, meaning that they have no ``missing baryons". Unexpectedly, the dynamics of our galaxies are dominated by the baryons, leaving small room for dark matter inside their discs.
%%%%%%%%%%%%%%%%%%%%%%%%%%%%%%%%%%%%%%%%%%%%%%%
\keywords{galaxies: dwarf --- galaxies: formation --- galaxies: evolution --- galaxies: kinematics and dynamics --- dark matter}
%% add here a maximum of 10 keywords, to be taken form the file <Keywords.txt>
%%%%%%%%%%%%%%%%%%%%%%%%%%%%%%%%%%%%%%%%%%%%%%%
\end{abstract}

\firstsection % do not remove

\section{Introduction}

H\,{\sc i}--kinematics play a major role in our understanding of galaxy formation and evolution. In particular, H\,{\sc i} rotation curves allow us to study the rotation of galaxies and the distribution of matter inside them (e.g. \citealt{deblok,swaters,noordermeer}). 

Despite this, H\,{\sc i} observations remain on very early stage in one of the most studied galaxy populations in the last years: ultra-diffuse galaxies (UDGs, \citealt{vandokkum}). While this long-ago discovered population (e.g. \citealt{impey}) has been characterized relatively well with deep imaging (e.g. \citealt{mihos,vanderburg,roman2,greco,paperII}, and references therein), most of what we know from their H\,{\sc i} kinematics come from a few, single-dish based, studies (e.g. \citealt{leisman,spekkens}). Expanding this line of investigation is essential to try to understand why these galaxies have effective radii of normal spiral galaxies but surface brightness of the faintest dwarfs, and if they show any other unusual property.

\citet{leisman} carried out the most systematic study of H\,{\sc i} in UDGs up to date, by studying the sources in the ALFALFA catalogue \citep{alfalfa} that meet the optical definition for being a UDG. Studying such sample, the authors realized that those galaxies have a global H\,{\sc i} profiles narrower than galaxies of similar masses, what may suggest that their H\,{\sc i}--rich UDGs rotate more slowly than expected. However, such profiles were not corrected for inclination, as the latter is usually unfeasible to estimate from optical images due to the low surface brightness nature of this galaxies, and in general global profiles are not as reliable as rotation curves. \citet{leisman} also had resolved H\,{\sc i} data for three galaxies, but 2D approaches in low-resolution data as theirs tend to lead to unreliable kinematics due to beam smearing (e.g. \citealt{bosma,barolo}). Given this, the question of whether or not UDGs have a different kinematic signature than other galaxies with similar masses remains open.

In order to give a more conclusive answer to such question, in this work we undertake 3D kinematic modeling, unaffected by beam smearing, of resolved H\,{\sc i} data of six H\,{\sc i}--rich UDGs. The talk on which this text is motivated is based on the work by \citet{btfr}.

\section{H\,{\sc i} and optical data}
Our H\,{\sc i} observations come from two radio telescopes, the Karl G. Jansky Very Large Array (VLA) and the Westerbork Synthesis Radio Telescope (WSRT). We have a typical spatial resolution of 2 independent beams per galaxy side, and a spectral resolution of $\approx$ 4 and 6 km~s$^{-1}$ for the VLA and WSRT data cubes, respectively. 

We complement this with deep optical imaging of our sources, obtained with the One Degree Imager on the 3.5-meter WIYN telescope at the Kitt Peak National Observatory. Specifically we observe our galaxies using the $g$ and $r$ filters with a total integration time of 45 minutes per filter.

Detailed information on the observing strategies, data reduction process and characterization of the data can be found in \citet{leisman} and Gault et al. in prep.\\

\noindent
Using the data cubes and the distances to the galaxies reported in \citet{leisman}, we can derive the H\,{\sc i} mass of our UDGs. This measurement is rather accurate: H\,{\sc i} fluxes can me measured with good precision from the data cube and the distances to our sample are distant enough (mean distance $\sim$ 90 Mpc) to be well represented by the Hubble flow distance with reasonably small uncertainties. We estimate the total mass in gas by correcting for the presence of helium, $\mathrm{M_{gas} = 1.33 \times M_{HI}}$. The stellar mass is obtained by means of the $\mathrm{M/L}$--color relation by \citet{herrmann}, with magnitudes measured from our WIYN images. Then, we combine the gass and stellar mass to derive the baryonic mass, which is mainly given by the gas mass (mean $\mathrm{M_{gas}/M_\star \approx 15}$), and therefore basically unaffected by any possible systematics while estimating the stellar mass.

\section{3D gas kinematics}
Given the low spatial resolution of our sample, rotation velocities derived with conventional 2D methods would be strongly affected (e.g. \citealt{bosma}). To mitigate this effect, we use the software $\mathrm{^{3D}Barolo}$ \citep{barolo}, which is largely unaffected by beam smearing since it fits tilted ring models directly to the data-cube instead of to the velocity field.

Since the rotation velocity and velocity dispersion will be left as a free parameters for the kinematic modelling of $\mathrm{^{3D}Barolo}$, we need to specify the position angle and inclination of the galaxies. The position angle is estimated by finding the orientation that maximizes the amplitude seen in the position--velocity diagram. The inclinations are derived by minimizing the residuals between the observed total moment map of each galaxy, and the total moment maps of models of the same galaxy projected at different inclinations between 10$^\circ$--80$^\circ$. We test this method in a sample of 32 dwarf rich galaxies drawn from the APOSTLE hydrodynamical simulations \citep{apostle2,apostle}, that are ``observed" at similar S/N and resolution as our data using the \textsc{martini} software (\citealt{kyle2019}, version 1.0.2). We find that we can recover the position angle and inclination within $\pm$ 8$^\circ$ and $\pm$ 5$^\circ$, respectively, as long as the inclination of the galaxy is larger than 30$^\circ$. These uncertainties do not produce a significant effect in the recovered rotation velocities.

\begin{figure}
    \centering
    \includegraphics[scale=0.46]{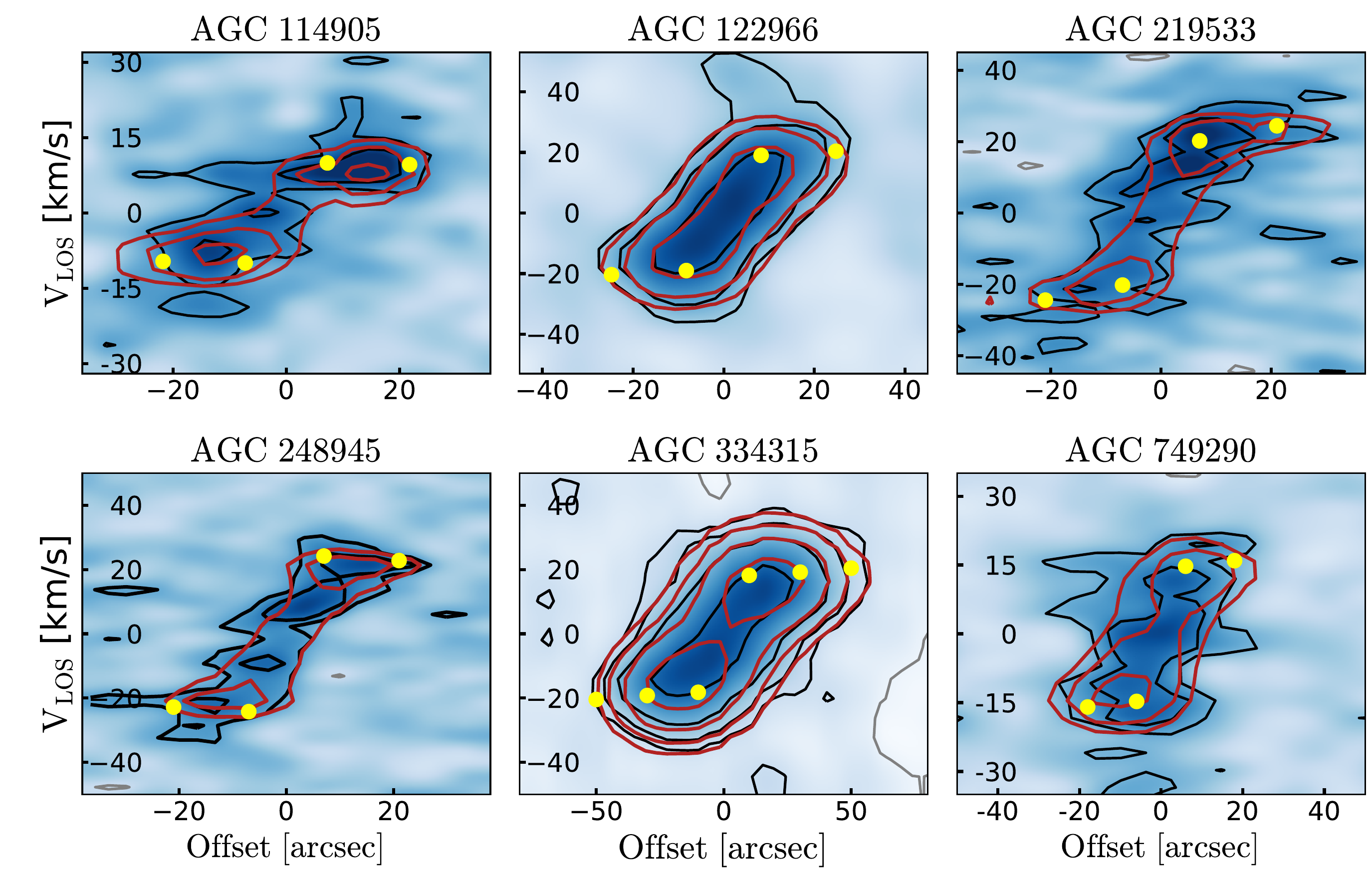}
    \caption{PV slices along the major axes of our  galaxies. Black and red contours show the data and best-fit tilted-rings model obtained with the software $\mathrm{^{3D}}$Barolo, respectively. The yellow points show the recovered rotation curves.}
    \label{fig:pvs}
\end{figure}

With the position angle and inclination fixed, we run $\mathrm{^{3D}Barolo}$ in our UDGs. For our six galaxies $\mathrm{^{3D}Barolo}$ converges and the best-fit model seems to well represent the data, as shown in Figure~\ref{fig:pvs}. We use $\mathrm{^{3D}Barolo}$ to convert our rotation velocities to circular speeds by means of the asymmetric drift correction (see \citealt{iorio}). This correction turns out to be very small for all the galaxies ($\leq$ 2~km~s$^{-1}$), mainly because the galaxies show low velocity dispersions, as it can be glimpsed from the narrowness of the position-velocity diagrams.

{}

\section{Results and discussion}
\subsection{UDGs in the $\mathrm{M_{bar}-V_{circ}}$ plane}
\begin{figure}
    \centering
    \includegraphics[scale=0.52]{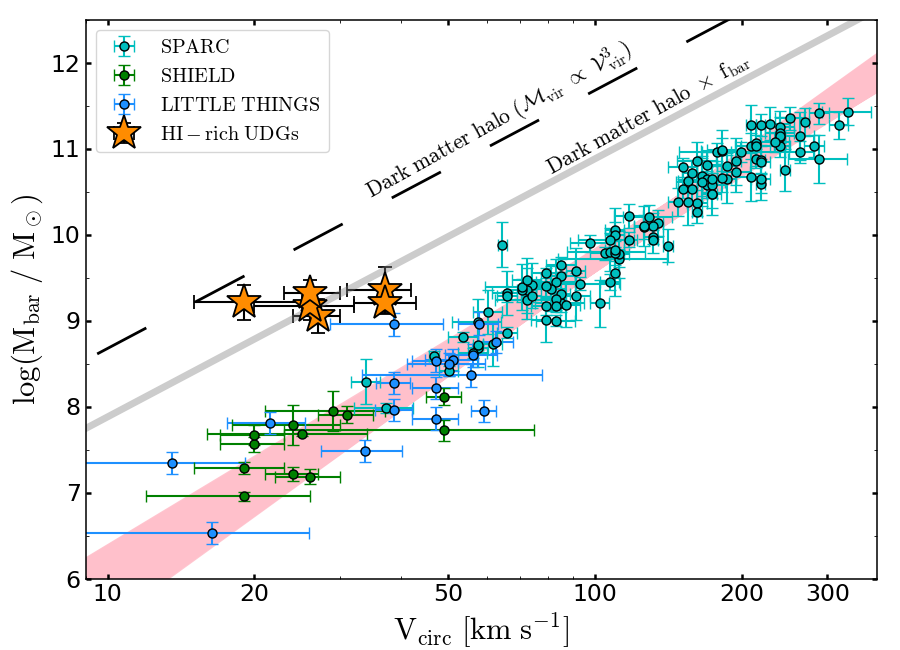}
    \caption{$\mathrm{M_{bar}-V_{circ}}$ plane. Galaxies from the SPARC, SHIELD and LITTLE THINGS lie on top of the BTFR. The pink area shows the 99\% confidence interval of an orthogonal distance regression to the SPARC sample. H\,{\sc i}--rich UDGs are outliers of the BTFR, in a position consistent with having no missing baryons. Taken from \citet{btfr}.}
    \label{fig:btfr}
\end{figure}

With the baryonic masses and circular velocities of our UDGs, we are in position to study the $\mathrm{M_{bar}-V_{circ}}$ plane, where galaxies follow remarkably tightly the baryonic Tully-Fisher relation (BTFR, e.g. \citealt{mcgaugh1,mcgaugh2}). In Figure~\ref{fig:btfr} we show our galaxies in such plane, together with galaxies from the SPARC \citep{sparc}, LITTLE THINGS \citep{iorio} and SHIELD \citep{shield} samples. While these last three samples follow the BTFR, our H\,{\sc i}--rich UDGs are clear outliers well above the relation, with circular velocities too low for their baryonic masses.

{}

Before discussing the possible implications of this result, one may wonder how robust are our measurements and if the position of our galaxies is compatible with the BTFR in any possible way. Here we briefly discuss why this is not the case and how possible systematics in the circular velocities or baryonic masses cannot solve the observed discrepancies:
\begin{itemize}
    \item \underline{Could $\mathrm{M_{bar}}$ be overestimated?} Given the small importance of the stellar mass, any possible systematic in the baryonic mass should be related with the H\,{\sc i} component, which depends on the flux and distance. The fluxes from H\,{\sc i} line observations can be measured with good accurancy, and we find results compatible with the ALFALFA single-dish observations \citep{leisman}. The distances come from the Hubble flow, and given the mean distance of $\sim$~90 Mpc, peculiar velocities can be neglected, making the systemic velocities of our galaxies a reliable measure of their distances. Therefore, an overestimation of a factor 10--100 is completely excluded.
    \item \underline{Could $\mathrm{V_{circ}}$ be underestimated?} The circular velocity depends mainly in the observed rotation velocity and the assumed inclination. As suggested by Figure~\ref{fig:pvs}, we find rotation velocities consistent with flat rotation curves; more importantly, our rotation curves are very extended, with an outermost radius $\sim 8-18$ kpc depending on the galaxy. Rotation velocities measured at such large radius are already tracing the flat part of the rotation curve for any plausible dwarf-galaxy dark matter halo (e.g. \citealt{kyle2015}, see their Fig.~2). Regarding the inclination, all our galaxies would need to be nearly face-on (inclination $\approx$ 10$^\circ$--20$^\circ$), which is very unlikely and incompatible with the observed H\,{\sc i} maps. Finally, it is worth stressing that non-circular motions cannot systematically bias the recovered circular velocities towards lower values \citep{kyle2019}. \\
\end{itemize}

Given all of the above, we conclude that the position of our H\,{\sc i}--rich UDGs are robust and indeed these galaxies do not follow the BTFR. This result is in agreement with the suggestion by \citet{leisman} of UDGs rotating more slowly that galaxies with similar mass. 

MOdified Newtonian Dynamics (MOND, e.g. \citealt{mcgaugh2}; \citealt{mond}) predicts that, if galaxies are in dynamical equilibrium and relatively well isolated, they should follow the relation $\mathrm{M_{bar} \propto V_{circ}^4}$. Our galaxies should have had enough time to be in equilibrium even if $all$ of them interacted with their nearest neighbor, and by selection are isolated. However, they do not follow the BTFR, and may present a challenge to MOND.

\subsection{The baryon fraction of H\,{\sc i}--rich UDGs: no ``missing baryons"}
While the result of the UDGs being off the BTFR is already surprising, their position in the $\mathrm{M_{bar}-V_{circ}}$ plane reveals something else. The black dotted line in Figure~\ref{fig:btfr} shows the relation between the circular velocity at the virial radius and the virial mass of a galaxy halo (dark matter plus baryons, $\mathrm{M_{vir}/M_\odot}~\approx~4.75~\times~10^{5}~\mathrm{(V_{vir}/km~s^{-1})^3}$, cf. \citealt{mcgaugh3}). If multiplied by the cosmological baryon fraction ($\mathrm{f_{bar}}~\approx$~0.16), it gives rise to a relation where galaxies with a baryon fraction equal to the cosmic mean should lie, indicated by the grey solid line. Unexpectedly, the position of our galaxies are consistent with this line, indicating that they have the cosmological baryon fraction, meaning that they are compatible with having no ``missing baryons".

This result is in principle counterintuitive: these galaxies are dwarfs, with relatively weak potential wells, and are mostly made of gas. How could they have retained all of their baryons? Our idea, motivated also by the currently weak gas heating indicated by the low velocity dispersions, is that feedback processes have been rather inefficient, and the galaxies have not ejected a significant amount of gas, or they have promptly re-accreted it. 

\subsection{Low dark matter content}
Since our galaxies seem to have more baryons than usual, and yet they rotate slowly compared with galaxies of similar masses, a natural question is how is the dark matter content of this galaxies. To study this we derive the dynamical mass of our galaxies, using the relation $\mathrm{M_{dyn}(R<R_{out}) =  V_{circ}^2~R_{out} / G}$, with $\mathrm{R_{out}}$ the radius of the outermost point of the rotation curve. Or sample has a mean $\mathrm{R_{out}/R_{\rm d}} = 4$, with $\mathrm{R_{\rm d}}$ the optical disc-scale length. Then, we compute the baryonic to dynamic mass ratio ($\mathrm{M_{bar}/M_{dyn}}$). We find that our galaxies have a ratio much higher than expected, very close to unity, indicating that the baryonic mass dominates the dynamics of these galaxies. In fact, our measurements imply that our galaxies have, inside their discs, dark matter fractions smaller than 0.5 and compatible with 0, meaning that they have little room, if any, for dark matter. This is contrary to what is observed in most low surface brightness galaxies, where the dark matter dominates at all radii.

Recently, based on the velocity dispersion of their globular clusters, the UDGs NGC1052-DF2 \citep{vandokkumDM, danieli} and NGC1052-DF4 \citep{vandokkumDM2} have been claimed to have none or little dark matter, but some concerns exist regarding their distances and environments \citep{trujilloDF2, monelli}. Our UDGs have robust distances from their recession velocities and are relatively isolated (mean distance to nearest neighbor $\sim$~1 Mpc), mitigating these concerns. Perhaps they could be subject to different systematics, but demonstrate that there may indeed exist a population of unusually dark matter-deficient galaxies.

The dynamical properties here shown, namely the shift from the BTFR and the low dark matter content are similar to those in tidal dwarf galaxies \citep{lelliTD}. Given the isolation and blue colors of our UDGs a possible tidal dwarf origin for all of them does not seem likely. However, testing further this hypothesis is hard with the current data. %Also, tidal dwarf galaxies are expected be more compact than normal dwarfs (see \citealt{tdgs_illustris}), while UDGs are selected for their large sizes. 

\end{document}